\documentclass[preprint,,showpacs,preprintnumbers,amsmath,amssymb]{revtex4-1}

\usepackage{graphicx}
\usepackage{epstopdf}
\usepackage{dcolumn}
\usepackage{bm}
\usepackage{amsmath}
\usepackage{graphicx} 
\usepackage{dcolumn}
\usepackage{bm}
\usepackage{amsmath}
\usepackage{epsfig}
\usepackage{float}
\usepackage{epstopdf}
\begin{document} 

\title{Light-Driven Raman Coherence as a Non-Thermal Route to Ultrafast Topology Switching 
}
	




\author
{C. Vaswani$^{1}$, L.-L. Wang$^{1}$, D. H. Mudiyanselage$^{1}$, Q. Li $^{2}$, P. M. Lozano $^{2}$, G. Gu$^{2}$, D. Cheng$^{1}$, B. Song$^{1}$, L. Luo$^{1}$, R.~H.~J.~Kim$^{1}$, C. Huang$^{1}$, Z. Liu $^{1}$, M. Mootz $^{3}$, I. E. Perakis$^{3}$, Y. Yao$^{1}$, K. M. Ho$^{1}$ and J. Wang$^{1\dag}$}

\affiliation{$^1$Department of Physics and Astronomy and Ames Laboratory-U.S. DOE, Iowa State University, Ames, Iowa 50011, USA. 
	\\$^2$Condensed Matter Physics and Materials Sciences Department, Brookhaven National Laboratory, Upton, NY 11973-5000, USA.
	\\$^3$Department of Physics, University of Alabama at Birmingham, Birmingham, AL 35294-1170, USA.}

\date{\today}

\begin{abstract}
A grand challenge underlies the entire field of {\em topology}-enabled quantum logic and information science: how to establish topological control principles driven by quantum coherence and understand the time-dependence of such periodic driving?
Here we demonstrate a THz pulse-induced phase transition in Dirac materials that is periodically driven by vibrational coherence due to excitation of the lowest Raman-active mode. 
Above a critical field threshold, there emerges a long-lived metastable phase with unique Raman coherent phonon-assisted switching dynamics, absent for optical pumping. The switching also manifest itself by 
non-thermal spectral shape, relaxation slowing down near the Lifshitz transition where the critical Dirac point (DP) occurs,
and diminishing signals at the same temperature that the Berry curvature induced Anomalous Hall Effect varnishes.
These results, together with first-principles modeling, identify a mode-selective Raman coupling that drives the system from strong to weak topological insulators, STI to WTI, with a Dirac semimetal phase established at a critical atomic displacement controlled by the phonon pumping. 
Harnessing of vibrational coherence can be extended to steer symmetry-breaking transitions, i.e., Dirac to Weyl ones, with implications on THz topological quantum gate and error correction applications. 

\end{abstract}

\maketitle
Dynamic driving by periodic lattice vibrations represents a powerful approach to manipulate topological band structures, in stark contrast to equilibrium tuning methods, e.g., temperature, chemical substitution and static strain/electric/magnetic fields \cite{Bas17,Li2013}.
{\em Ultrafast non-thermal} manipulation of topology \cite{Li2, LLuo}, particularly at preferred terahertz (THz)-cycle clock rates, is key for full implementation of dynamical protocols needed to both match current information and sensing 
technologies and exceed their limits via topological functionalities \cite{lightwave, rohit, kemper,kira}. 
Despite of intriguing studies recently \cite{Fausti2011Science, MKoz19,Nelson19,Sie19}, topological states driven by THz optical phonons have not been explored, especially in Dirac semimetals.   
ZrTe$_5$ is a model Dirac system \cite{Li16, Wen14,Tan19} for establishing such topological quantum switching by periodic driving from phonons because of the minimal single nodal (Dirac) point and extreme sensitivity on the small structural changes across a broad range of phases, from STI to Dirac semimetal to WTI. However, only thermal- or strain-induced transition \cite{Xu18, Mut19} have been studied.   

We implement a dynamical topology-switching scheme using intense THz laser pulses (red line) to excite a Raman-active ($A_{1g}$) optical phonon mode, as illustrated in Fig. 1a. The subpicosecond THz driving has near-single-cycle electric field profile in the time domain and broadband spectrum with central frequency $\sim$1.2THz (gray shade, Fig. 1b). The atomic displacement associated with the $A_{1g}$ eigen-mode (Fig. 1c) is a translational rigid-chain structure which mostly involves opposite displacements along the b-axis of the dimer Te (Te$_d$), apical Te (Te$_a$) and Zr atoms along the b-axis arising from the neighboring Zr-Te units.  This motion results in a modulation of the atomic positions along b-axis that determines the van der Waals coupling and, in turn, controls the band inversion at the $\Gamma$ point. Most intriguingly, this mode could enable an exclusive topology switch without symmetry breaking because it has the extreme sensitivity to topology while preserves the inversion symmetry of the lattice. 
Coherent excitation of the $A_{1g}$ phonon expects to create a periodically-driven state via lattice vibrations that modulate the topological bands and switch from STI (top, Fig. 1a) to Dirac semimetal (middle) to WTI (bottom) phases.
Dephasing of this topological coherent state, via Dirac fermion-phonon interaction, leads to conversion of Raman phonon coherence into population, i.e., into finite atomic displacement 
associated with the establishment of a final state with highly non-thermal characteristics. 
However, the salient spectroscopy features for the ultrafast non-thermal phase transition in any Dirac materials were not established until now.

\begin{figure}
	\begin{center}
		\includegraphics[scale=0.6]{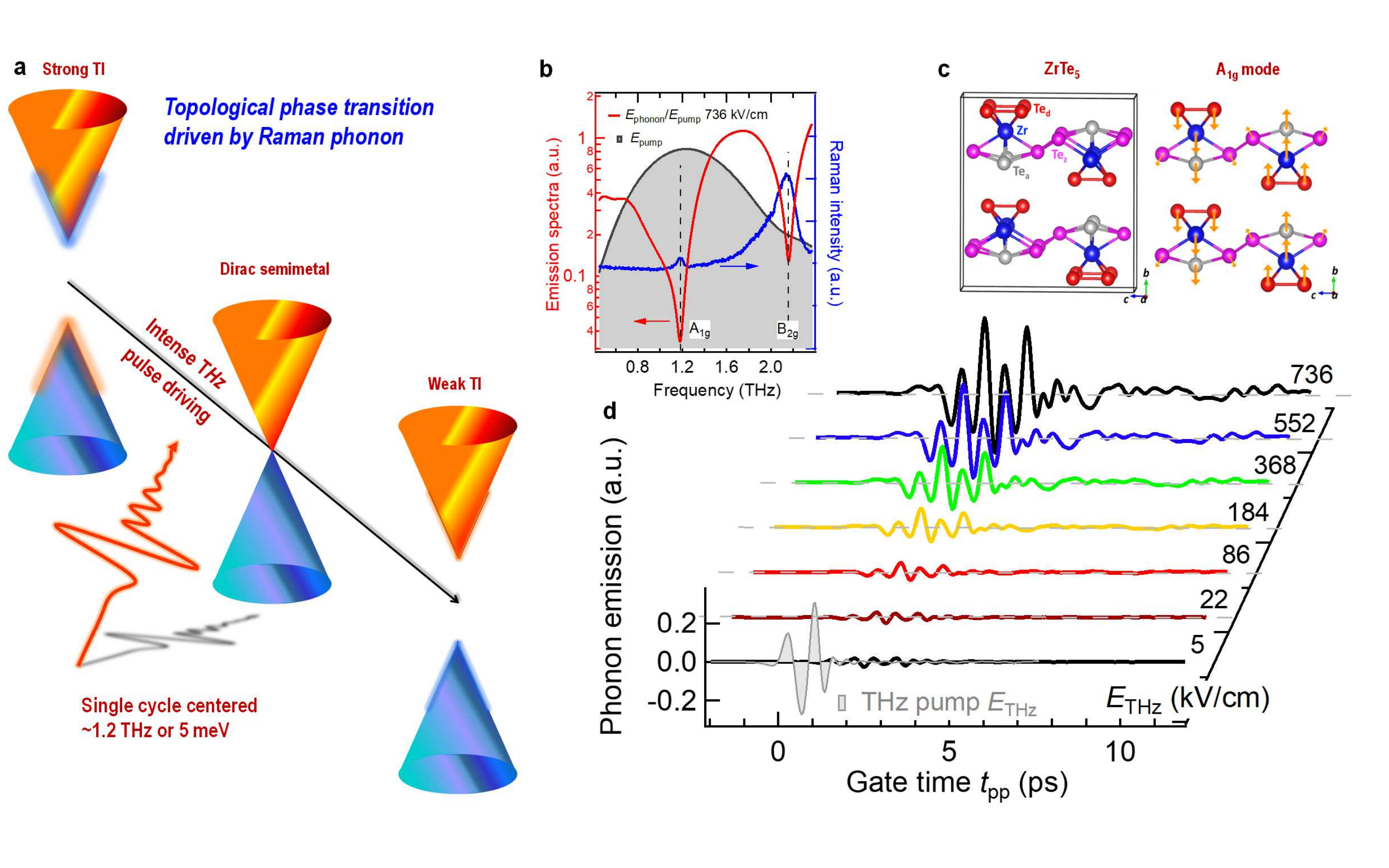}
	\end{center}	
	\textbf{Fig. 1.} \textbf{Intense THz-driven Raman phonon coherence in ZrTe$_5$.} (a) Schematic of the topological switching driven by the THz coherent excitation in Dirac semimetals. (b) Spectrum of the coherent phonon emission (red) obtained from the time domain traces shown for maximum THz field strengths E$_{max}$=736kV/cm. The pump THz spectrum is shown in gray. Dash lines indicate the identification of the excited modes with the measured Raman spectrum (blue).
	(c) Crystal structure of ZrTe$_5$ and the atomic displacements of the A$_{1g}$ Raman mode leading to the modulation of interlayer spacing. The arrows indicate the direction and magnitude of the atomic displacements. (d) Time-domain traces of the measured coherent phonon emission for various THz field strengths. The THz driving pulse is shown in gray. 	
	\label{Fig1}
\end{figure}


In this paper, we provide evidence for the distinct topology switching in ZrTe$_5$ driven by THz Raman phonon coherence.  
Our results are consistent with the calculations of $A_{1g}$ mode-selective electronic band structures. 
We used a bulk single crystal sample that exhibits a 3D linear dispersion and bulk bandgap less than 30 meV \cite{chen1, chen2, Mar19}. 
Coherent phonon emissions after intense THz pump excitation at 4.1K for various field strengths E$_{\mathrm{THz}}$=5, 22, 86, 184, 386, 552 and 736kVcm$^{-1}$ are sampled in the time domain by a weak optical pulse (Methods), as shown in Fig 1d.
A pronounced multi-cycle oscillation is clearly visible in the sample emission. The Fourier transform (FT) spectra of these coherent beatings at 736kVcm$^{-1}$ display two dominant peaks centered at $\sim$1.2THz and 2.1THz. The static Raman spectrum (blue line, Fig. 1b) from the same sample, shown together, identifies their Raman symmetry (dash lines). Note that the strongest emission peak $\sim$1.2THz matches very well with the A$_{1g}$ mode, unlike for the B$_{2g}$ mode dominant in the static Raman spectra. 
This observation clearly shows that the intense THz driving strongly excites a Raman A$_{1g}$ coherence 
in the driven state, unlike for the equilibrium state, which periodically modulates the interlayer spacing. 
In addition, the three infrared (IR) active phonon modes with frequencies 0.63THz, 1.5THz and 2.3THz, as seen in linear THz transmission (Fig. S2, supplementary) are negligibly small in the intense THz-driven state (Fig. 1d), i.e., the preferred coupling of the topological electronic bands to the $A_{1g}$ mode. 
This Raman mode-selective coupling attests the excitation scheme (Fig. 1c) due to the extreme sensitivity of the band inversion to the interlayer spacing, which is reproduced by density functional simulations below. 

To characterize the observed periodically-driven non-equilibrium topological state dressed by Raman coherence Fig. 2a plots the THz differential transmission change $\Delta E/E_{0}$ (red circles) after excitation by the intense THz pump pulse (gray shade) as a function of the pump–probe time delay $\Delta t_{pp}$ at 4.1K. The measured $\Delta E/E_{0}$ signal at gating pulse delay t$_{gate}$=0ps (inset) originates mainly from phonon renormalization and recovery induced by Dirac fermion-phonon ($e$-$ph$) interaction (further discussed in Fig. 3a). 
Shown together is the pump-induced coherent phonon emission (black line, overlaid) that is measured simultaneously as a function of $\Delta t_{pp}$. 
These results reveal the build-up of a metastable state, which occurs exclusively during the coherent Raman phonon oscillations in time. 
At the early times during the THz pulse, marked by t$_{pulse}$ (Fig. 2a), there is only small pump-induced $\Delta E/E_{0}$ signals. 
This excludes the THz heating of electronic states near the Fermi surface, which would lead to quasi-instantaneous increase in the transient signals on the arrival of the pump pulse. 
In contrast, the transient state evolution, seen from the pronounced $\Delta E/E_{0}$ signals, only occurs at later times after the pulse, but before phonon dephasing, marked as t$_{dephasing}$, i.e. during the period of the pronounced coherent phonon vibrations (blue line, Fig. 2a). 
This emergent behavior after the incident THz excitation dominates the driven state dynamics. The formation process is followed by a quasi-steady temporal regime that marks the establishment of a final metastable state after the dephasing of the Raman phonon coherence by, e.g., the strong $e$-$ph$ couping.  
In contrast, the build-up behavior during coherent oscillations is absent for high-frequency pump pulses tuned at 1.55 eV (Fig. 2b). 
Here we only see a sub-ps rise which now occurs mostly during the photoexcitation. This stark difference clearly indicates the non-thermal nature of the THz-driven state mediated by coherent Raman phonons.

\begin{figure}
	\begin{center}
		\includegraphics[scale=0.6]{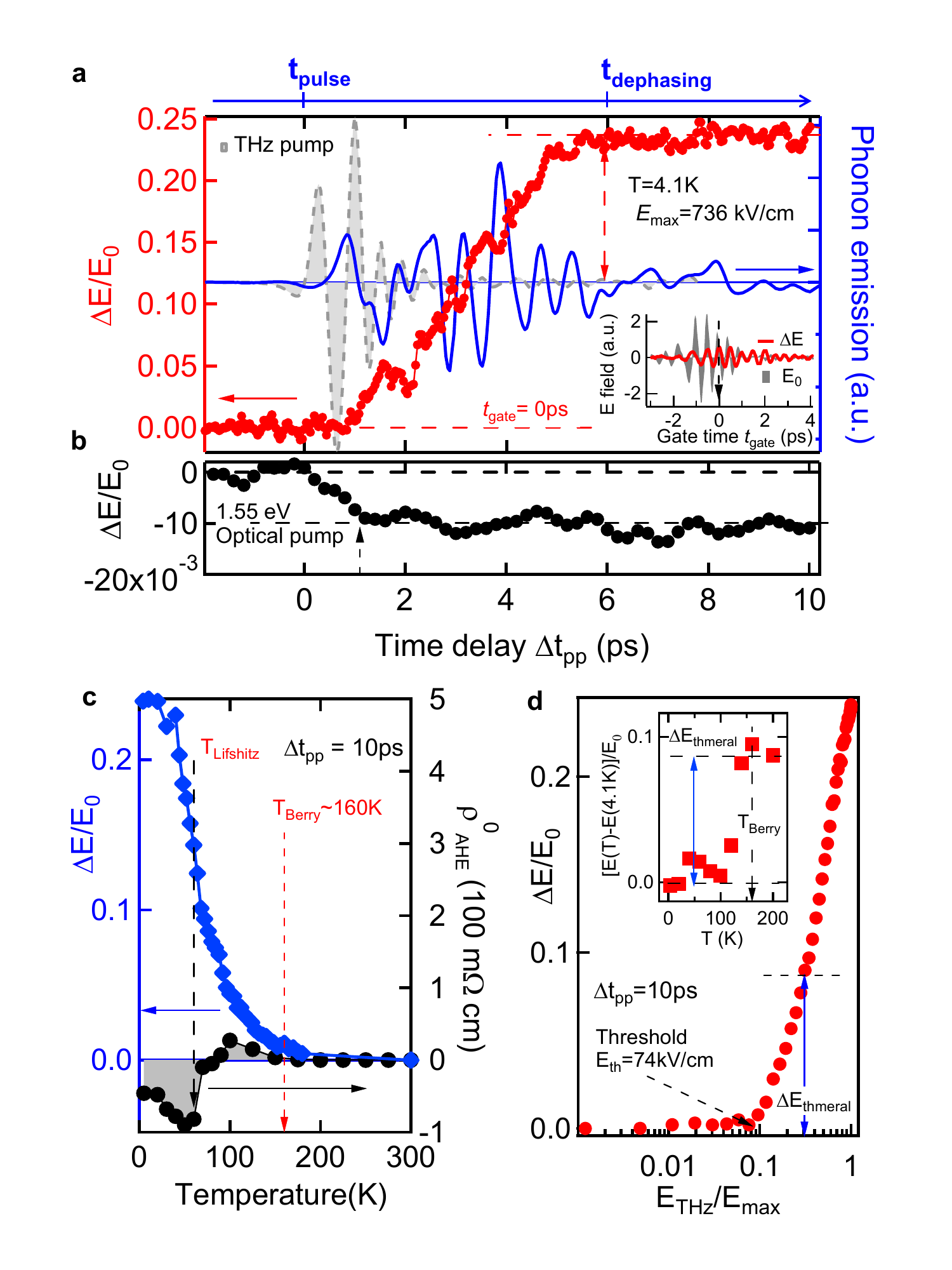}
	\end{center}	
	\textbf{Fig. 2.} \textbf{Light-driven topological phase transition above threshold THz pump field and at low temperature that the Berry curvature dominates.}	
	(a) THz differential transmission ($\Delta E/E_0$, red circles) as a function of the pump-probe time delay ($\Delta$t$_{pp}$), for a peak pump E-field of 736kV/cm. The blue trace shows the simultaneously measured phonon emission. The THz pump trace is shown in gray. The THz differential transmission starts to build-up during the coherent phonon emission at much longer times than the pump THz pulse. 
	Inset: time domain THz raw data. The transmitted E field through sample E$_{0}$ (gray shade) and the pump induced transmitted probe E field change $\Delta$E (red line) are measured by scanning t$_{gate}$. (b) THz differential transmission for the case of high energy optical pump(1.55eV) shows a much faster rise time. 
	(c) Temperature dependence of the differential THz transmission in the metastable state ($\Delta$t$_{pp}$=10ps, blue diamonds) indicates strong correlation with the Berry curvature-induced Anomalous Hall effect (black solid circles).   
	(d) Pump E-field dependence of the THz differential transmission for THz pump at ($\Delta$t$_{pp}$=10ps), indicating a threshold pump field for the formation of the metastable state. Inset: The thermal topological change, $\Delta_{thermal}$, required for the thermally driven STI-DP-WTI transition is determined by the change of THz field transmission between 4.1K and 160K (T$_{Berry}$).  
	\label{Fig2}
\end{figure}

Experimental evidence associating the observed phase evolution with the topological switching is presented in Figs. 2c and 2d as follows. The first evidence is to compare the temperature dependence of the pump-probe $\Delta E/E_{0}$ signals in the meta-stable states (at $\Delta$t$_{pp}=10ps$) with that of the Anomalous Hall Effect (AHE) that directly probes the Berry curvature $\Omega _{k}$ generated by the Weyl nodes. As shown in Fig. 2c, the nonlinear THz signal (blue diamond) quickly diminishes at the same temperature, T$_{Berry}\sim$160K, where AHE Resistivity $\rho_{AHE}^{0}$ vanishes (black circles). 
In the AHE measurement, application of a magnetic field transforms a Dirac semimetal into a Weyl semimetal by breaking time reversal symmetry.
Consequently, Weyl nodes behave like magnetic monopoles that generate large Berry curvatures and act like an effective magnetic field. This gives rise to a non-zero $\rho_{AHE}^{} =\Omega _{k}\times E$. 
Here we obtained the temperature dependence of $\rho_{AHE}^{0}$, i.e., the saturation value of $\rho_{AHE}(B)$ (Fig. S1, Supplementary materials), by subtracting the ordinary Hall signals (linear background) at high magnetic field from the experimentally measured Hall resistivity \cite{liang2018}. 
It is clearly visible that $\rho_{AHE}^{0}$ in ZrTe$_5$ emerges below T$_{Berry}$ when the dominant carriers are Dirac fermions with linear dispersion near the conical point with conserved chirality. 
The T$_{Berry}$ correlates very well with the critical temperature associated with the THz-driven metastable phase transition (blue diamond, Fig. 2c). Therefore, the metastable phase has the same topological origin as the chiral magneto-transport and cannot be established by excitation of normal Fermi surface dominant above T$_{Berry}$.  
Note also that the sign change of $\rho_{AHE}^{0}$ in the vicinity of T$_{Lifshitz}\sim$60K where the critical DP occurs separating the STI-WTI transitions\cite{Xu18} (Fig. S1b, supplementary), agrees with the rapid rise of the pump-probe signals, marked by the black dash line, associated with the metastable state. 

The second evidence is a distinct nonlinear pump fluence dependence of the pump-probe signals with a larger size than the change, $\Delta_{thermal}$, required for the thermally-driven topological transition. 
By increasing temperatures from 4.1K to 160K (T$_{Berry}$), as shown in the inset of Fig. 2d, the $\Delta_{thermal}$ can be directly determined $\sim$0.09 corresponding to the change of THz field transmission during the STI-DP-WTI transition.   
Here we compare the $\Delta_{thermal}$ with the THz pump field dependence of the differential transmission $\Delta E/E_{0}$ signals at a fixed time $\Delta t_{pp}=$10ps (Fig. 2d). We emphasize two key points. First, it is clearly visible that the pump-induced $\Delta E/E_{0}$ is negligibly small at THz field strengths less than E$_{\mathrm{th}}\sim75$kV/cm. This threshold behavior of the formation dynamics (Fig. 2a) is not limited by our noise floor, which is a hallmark of the non-equilibrium phase transition to a THz-driven metastable state. Second, at slightly higher field above E$_{\mathrm{th}}$, the pump-induced differential transmission surpasses the $\Delta_{thermal}$ value for the thermally-induced STI-DP-WTI topological switching (black dash line).  
This indicates that the sufficiently large lattice displacement above E$_{\mathrm{THz}}$ drives the system cross the topological phase boundary to new band structures determined by phonons.

\begin{figure}
	\begin{center}
		\includegraphics[scale=0.6]{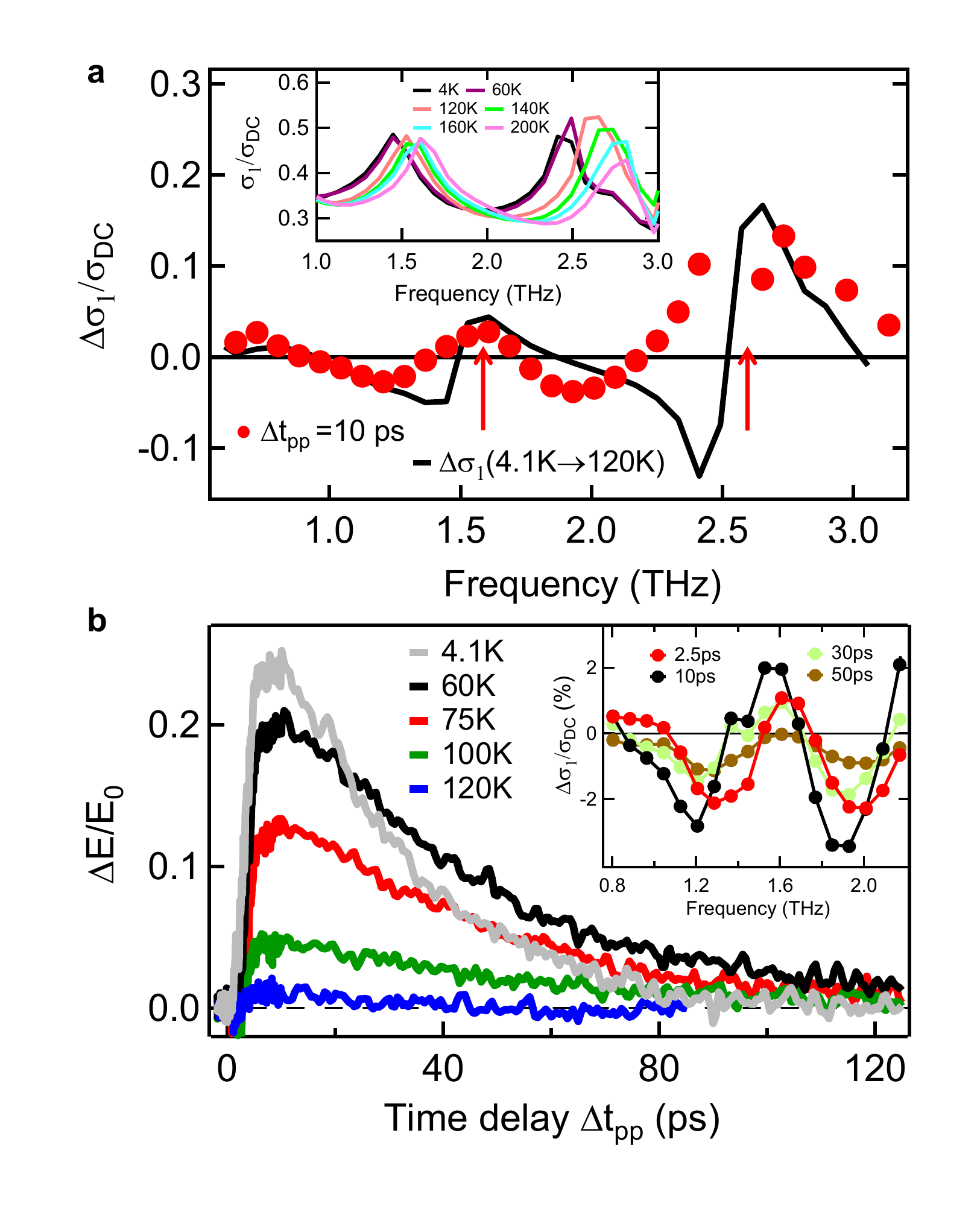}
	\end{center}
	\textbf{Fig. 3.} \textbf{The distinct spectral and temporal features of the non-thermal, metastable states.} (a)  Pump-induced change in conductivity ($\Delta \sigma_{1}/\sigma_{DC}$, red circles) at T=4.1K compared with the thermal conductivity change (black circles) between 120K and 4.1K, i.e., STI and WTI states, at $\Delta$t$_{pp}$=10ps. Inset shows the shift in the IR phonon frequencies in the static conductivity along the a-axis  as a function of temperature. 	
	(b)	Differential THz transmission profiles ($\Delta E/E_0$) for various temperatures show that the relaxation of the metastable state is the longest near T$_{Lifshitz}$=60K. Inset: THz response $\Delta \sigma_{1}/\sigma_{DC}$ as a function of frequency for time delays $\Delta$t$_{pp}$=2.5ps, 10ps, 30ps and 50ps
	at E$_{THz}$=736kV/cm at 4.1K.
	\label{Fig3}
\end{figure}

Next we identify further some distinguishing spectral and temporal features associated with the THz-driven metastable phase that are different from the thermal states. First, Fig. 3a reveals a distinct, non-thermal, spectral shape in the non-equilibrium response function. 
At equilibrium, the frequency-dependent conductivity $\sigma_{1}/\sigma_{DC}$  (inset, Fig. 3a) reveals two strong IR-active phonons along the a-axis (probe direction) with resonant peaks at $\omega_{IR}^{1,2}\sim$1.5 and 2.5THz in the 4.1K trace (red line). These IR phonon modes can contribute to the the Raman A$_{1g}$ phonon generation via the ionic Raman mechanism that involves the IR modes as mediator and IR-Raman coupling due to anharmonicity \cite{Raman}. 
At elevated temperatures, these modes progressively shift to higher frequencies up to 200K (magenta line). In the THz-driven phase, the pump-induced conductivity change $\Delta\sigma_{1}/\sigma_{DC}$ at $\Delta t_{pp}=$10 ps shows spectral oscillations(red circles, Fig. 3a) with pronounced absorptive features (red arrows).  In contrast, the normal state thermalization leads to dominantly inductive spectral shape (black line, Fig. 3a) which can be obtained by subtracting $\sigma_{1}$ traces at higher temperature and 4.1K, i.e., between STI and WTI thermal states. This result highlights the difference between the driven phase evolution and temperature-/laser-heating induced phase thermalization process. 
Second, Fig. 3b plots the relaxation dynamics that measures the lifetime of the THz-driven state from 4.1K to 120K. 
The temporal profile is consistent with the slow build-up, peaked at $\sim$10ps, and $\sim$100ps decay of phonon renormalization as seen in the $\sigma_{1}/\sigma_{DC}$ spectra shown for various $\Delta t_{pp}$ delays (inset, Fig. 3b). 
The transient phase decays with a single exponential profile over $\sim120$ps, see e.g. the 60K trace (black line). Such relaxation is nearly 2 orders of magnitude longer than that reported for the case of optical excitation using higher energy photons \cite{Man15, Zha19}, due in part to the minimal heating of the Fermi surface due to the THz pumping.
Most intriguingly, the relaxation time exhibits a non-monotonic temperature dependence. It firsts becomes longer with temperature increase from 4.1K (gray line), reaches a maximum at $\sim$60K (black line), and finally decreases with temperature up to 120K (blue line), as shown in Fig. 3a. Critical to note that the longest lifetime appears $\sim$T$_{Lifshitz}$ at which the critical Dirac point appears \cite{Xu18}. 
This dynamical slowing down further underscores the topological origin of the THz-driven phase transition. 

\begin{figure}
	\begin{center}
		\includegraphics[scale=0.6]{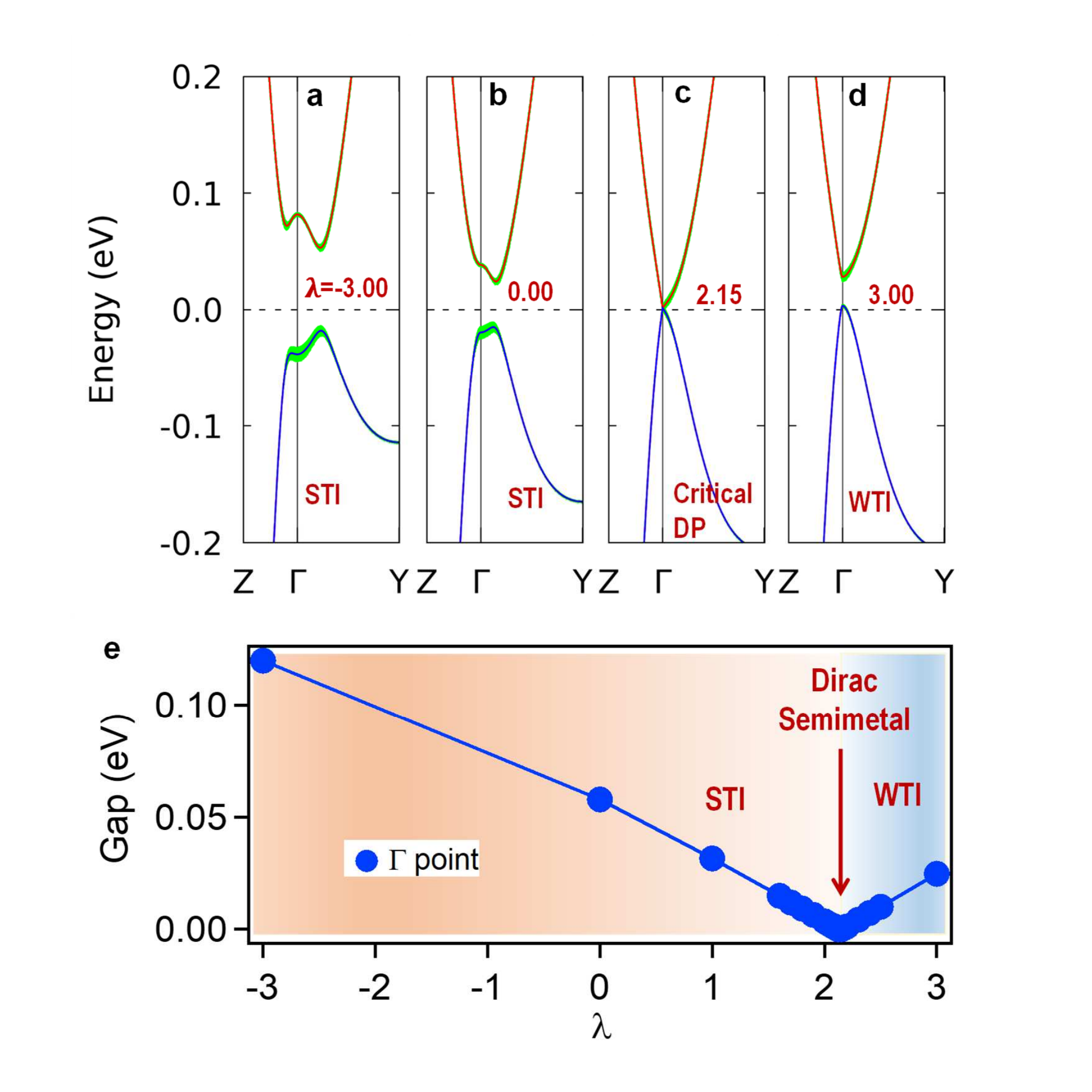}
	\end{center}
	\textbf{Fig. 4.} \textbf{Phonon and band structure calculations for the topological phase switching via excitation of the A$_{1g}$ phonon mode in a Dirac semimetal ZrTe$_5$.} 
	The band structure along Z-$\Gamma$-Y for different magnitudes of the atomic displacement ($\lambda$) corresponding to the A$_{1g}$ mode: (a) -3.00 (b) 0.00, (c) 2.15 and (d) 3.00. The green shadow indicates the projection of Te$_d$ p orbitals denoting band inversion. This procedure identifies the states with $\lambda=-3.00,0.00$ (a), (b) as STI having a topological index of (1;110) whereas the critical Dirac point occurs for $\lambda =2.15$ (c). Absence of band inversion for $\lambda =3.00$ (d) indicates a WTI with the topological index (0;110). (e) Energy gap at $\Gamma$ point as calculated from the band structure as a function of atomic displacement magnitudes $\lambda$.	
	\label{Fig4}
\end{figure}

To put the observed Raman mode-selective topological phase transition into a stronger footing, we have carried out phonon and band structure calculations based on density functional theory (DFT) that provide theoretical support for the topological switching due to the coherent A$_{1g}$ phonon driving. 
Among the calculated zone-center optical phonon modes with the relaxed unit cell (Methods), the lowest Raman A$_{1g}$ mode of 1.12 THz and B$_{2g}$ mode of 2.18 THz compare well with the experimental values\cite{Lan84} of 1.18 THz and 2.16 THz, respectively. To study the effect of this A$_{1g}$ mode on the electronic structure of ZrTe$_5$, we vary the displacements ($\lambda$) from the experimental atomic positions and calculate the new band structure. For $\lambda$=1.0, the corresponding atomic displacements are 0.033$\AA$ for Zr, Te$_d$ and Te$_a$, and 0.017$\AA$ for Te$_z$. The band structure of ZrTe$_5$ along Z-$\Gamma$-Y for different displacements $\lambda$ is shown in Fig. 4. First, without any atomic displacements from equilibrium ($\lambda$=0.0) (Fig.4a), ZrTe$_5$ has a narrow band gap of 0.04 eV. Both the valence and conduction band edges appear along $\Gamma$-Y and are off the $\Gamma$ point, towards the Y point. The projection (green shadow) on Te$_d$ p orbitals clearly shows the band inversion between the valence and conduction bands, which agrees with the gapped massless Dirac state observed in experiments \cite{Li16}. 
As one of the indicators, the 2D topological index on the k$_z$=0 plane is 1 (Fig.S3a), shown by the odd number of crossings for the Wannier charge centers (WCCs) moving along k$_y$. The overall topological invariant index is (1;110) for the initial state. 
For $\lambda$=–3.0 (Fig.4b), the band gap increases to 0.07 eV but the band inversion along Z-$\Gamma$-Y directions remains. 
The system is still a gapped massless (Dirac) state \cite{Tabert}. Most interestingly, in contrast to the above, by moving in the other direction with more positive $\lambda$, the band gap decreases. The valence and conduction bands touch at the $\Gamma$ point when $\lambda$=2.15 (Fig.4c). 
Then the band gap reopens as $\lambda$ increases further. For $\lambda$=3.0 (Fig.4d), the re-opened band gap has no band inversion along Z-$\Gamma$-Y, as seen from the orbital projection of the Te$_d$ p orbitals (inset, green line). The corresponding 2D topological index on the k$_z$=0 plane is now 0, shown by the even number of crossings for the WCCs moving along k$_y$ in Fig.S3b. The overall topological index becomes (0;110) for a WTI. 
Thus, with the coherently excited Raman A$_{1g}$ mode within one THz pulse cycle above threshold E$_\mathrm{th}$, ZrTe$_5$ can be driven into coherent topological oscillations between STI and WTI states, with an interesting critical Dirac point in between. 
The dephasing of the phonon vibration, from multiple Dirac fermion-phonon scattering and/or disorder effects, results in non-thermal phonon populations, which lead to a renormalized spectral shape different from thermal ones shown in Fig. 3a. 
This critical bulk DP point (Method) is marked by the dashline in the $\lambda$ dependence of the band gap (Fig 4e), which can exist at the $\Gamma$ point in ZrTe$_5$ as the phase boundary between gapped Dirac state and WTI. Such transition can be driven by the A$_{1g}$ Raman phonon mode consistent with the THz tuning experiment. 
       
In summary, we identify a previously-inaccessible tuning scheme via mode-selective Raman coherence that can control the band topology. We demonstrate THz-driven topological phase transition during coherent lattice oscillations in a Dirac material. Harnessing the resonant THz-driven coherence of specifically tailored modes with an intense THz pulse electric field may become a universal principle for steering other symmetry-breaking transitions to Weyl states or phase.  


	
	\section*{Acknowledgements}
	This work was supported by the U.S. Department of
	Energy, Office of Basic Energy Science, Division of
	Materials Sciences and Engineering (Ames Laboratory is
	operated for the U.S. Department of Energy by Iowa State
	University under Contract No. DE-AC02-07CH11358) (THz spectroscopy and DFT simulations).
	Sample development and transport measurements in Brookhaven National Laboratory (QL, PML, GG) were supported by the US Department of Energy, Office of Basic Energy Science, Materials Sciences and Engineering Division, under contract NO. DE-SC0012704.
	L.-L. W was supported by the Center for the Advancement of Topological Semimetals, an Energy Frontier Research Center funded by the U.S. DOE, Office of Basic Energy Sciences.
	M. M and I.E.P at the University of Alabama, Birmingham was supported by the US Department of
	Energy, Office of Science, Basic Energy Sciences under award no. DE-SC0019137.

\end{document}